\documentclass[oldversion]{aa}  
\usepackage{graphics,graphicx}
\usepackage{txfonts}
\usepackage{natbib}
\bibpunct{(}{)}{;}{a}{}{,}

\def\ch{{\it Chandra}}

\def\inte{{\it INTEGRAL}}
\def\epi{EPIC-pn}
\def\fay{Fe$_2$SiO$_4$}
\def\irsul{FeSO$_4$}
\def\hem{Fe$_2$O$_3$}
\def\lep{FeO(OH)}
\def\mg{MgSiO$_3$}
\def\xmm{XMM-{\it Newton}}

\def\kms{km\,s$^{-1}$}
\def\Halpha{\ifmmode {\rm H}\alpha \else H$\alpha$\fi}
\def\Hbeta{\ifmmode {\rm H}\beta \else H$\beta$\fi}
\def\Hgamma{\ifmmode {\rm H}\gamma \else H$\gamma$\fi}
\def\Hdelta{\ifmmode {\rm H}\delta \else H$\delta$\fi}
\def\Lya{\ifmmode {\rm Ly}\alpha \else Ly$\alpha$\fi}
\def\Lyb{\ifmmode {\rm Ly}\beta \else Ly$\beta$\fi}
\def\Lyg{\ifmmode {\rm Ly}\beta \else Ly$\gamma$\fi}
\def\fei{Fe\,{\sc i}}
\def\feii{Fe\,{\sc ii}}

\def\feiv{Fe\,{\sc iv}}

\def\hi{H\,{\sc i}}
\def\hii{H\,{\sc ii}}

\def\ciii{\ifmmode {\rm C}\,{\sc iii} \else C\,{\sc iii}\fi}
\def\civ{\ifmmode {\rm C}\,{\sc iv} \else C\,{\sc iv}\fi}
\def\cv{\ifmmode {\rm C}\,{\sc v} \else C\,{\sc v}\fi}
\def\cvi{\ifmmode {\rm C}\,{\sc vi} \else C\,{\sc vi}\fi}

\def\nvi{N\,{\sc vi}}
\def\nvii{N\,{\sc vii}}
\def\oi{O\,{\sc i}}
\def\oii{O\,{\sc ii}}
\def\oiii{O\,{\sc iii}}
\def\o5007{[O\,{\sc iii}]\,$\lambda5007$}
\def\oiv{O\,{\sc iv}}
\def\ov{O\,{\sc v}}
\def\ovi{O\,{\sc vi}}
\def\ovii{O\,{\sc vii}}
\def\oviii{O\,{\sc viii}}

\def\neix{Ne\,{\sc ix}}
\def\nex{Ne\,{\sc x}}

\def\fei{Fe\,{\sc i}}
\def\feii{Fe\,{\sc ii}}

\newcommand{\uu}{4U\,1820-30}

\def\o{\o}
\def\ltsim{\raisebox{-.5ex}{$\;\stackrel{<}{\sim}\;$}}

\begin{document}
   \title{XMM-{\it Newton} observation of 4U~1820-30:}

   \subtitle{Broad band spectrum and the contribution of the cold interstellar medium}

   \author{E.~Costantini
          \inst{1}
          \and
          C.~Pinto\inst{1}
		  \and 
		  J.S.~Kaastra\inst{1,2}
		  \and
		  J.J.M.~in't Zand\inst{1}
		  \and
		  M.J.~Freyberg\inst{3}
		  \and
		  L.~Kuiper\inst{1}
		  \and
		  M.~M\'{e}ndez
		  \inst{4}
		  \and
		  C.P.~ de Vries\inst{1}
		  \and
		  L.B.F.M.~Waters\inst{1,5}
          }

   \institute{SRON, Netherlands Institute for Space Research, Sorbonnelaan, 2, 3584\,CA, Utrecht, The Netherlands \\
              \email{e.costantini@sron.nl}
         \and
  		Astronomical Institute, Utrecht University, PO Box 80000, 3508\,TA Utrecht, The Netherlands
		\and
		Max-Planck-Institut %fuer extraterrestrische Physik, Giessenbach Str. 1, D-85740 Garching bei M\"unchen, 
		f\"ur extraterrestrische Physik, Giessenbachstr.\ 1, D-85748 Garching bei M\"unchen, Germany
          \and
		  Kapteyn Astronomical Institute, University of Groningen, Postbus 800, 9700 AV, Groningen, The Netherlands 
		   \and
		   Sterrenkundig Instituut Anton Pannekoek, University of Amsterdam, Science Park 904, P.O. Box 94249, 1090 GE Amsterdam, The Netherlands
		     }

   \date{Received/Accepted}

\authorrunning{E.~Costantini et al.}
\titlerunning{XMM-{\it Newton} observation of 4U~1820-30}

% \abstract{}{}{}{}{} 
% 5 {} token are mandatory
 
  \abstract
   {We present the analysis of the bright X-ray binary 4U~1820-30, based mainly on \xmm-RGS data, but using complementary data from \xmm-\epi, \inte, and \ch-HETG, to
   investigate different aspects of the source. The broad band continuum is well fitted by a classical combination of black body and Comptonized emission. The continuum shape and the high
   flux of the source ($L/L_{\rm Edd}\sim0.16$) are consistent with a "high state" of the source. We do not find significant evidence of iron emission at
   energies $\geq6.4$\,keV. The soft X-ray spectrum contain a number of absorption features. Here we focus on the cold-mildly ionized gas. The neutral gas column density is $N_{\rm H}\sim
   1.63\times10^{21}$\,cm$^{-2}$. The detailed study of the oxygen and
   iron edge reveals that those elements are depleted, defined here as the ratio between dust and the total ISM cold phase, 
   by a factor $0.20\pm0.02$ and $0.87\pm0.14$, respectively. Using the available dust models, the best fit points to a major contribution of Mg-rich
   silicates, with metallic iron inclusion. Although we find that a
   large fraction of Fe is in dust form, the fit shows that Fe-rich silicates
   are disfavored. The measured Mg:Fe ratio is $2.0\pm0.3$. Interestingly, this modeling
   may point to a well studied dust constituent (GEMS), sometimes
   proposed as a silicate constituent in our Galaxy. Oxygen and
   iron are found to be slightly over- and under-abundant, respectively (1.23 and 0.85
   times the solar value) along this line of sight. We also report the
   detection of two absorption lines, tentatively identified as part of an outflow of mildly ionized gas
   ($\xi\sim-0.5$) at a velocity of $\sim1200$\,\kms.}
  % conclusions heading (optional), leave it empty if necessary 
   {}


   \maketitle
%
%________________________________________________________________

\section{Introduction}\label{par:intro}

The interstellar medium (ISM) in the plane of our Galaxy is a dynamic and
complex environment, composed of mainly neutral matter in both gas and dust form and by a warmer gas phase in the form of diffuse emission in, and above, the Galactic plane. 
The properties of the cold phase in the diffuse ISM have been extensively studied at long wavelengths, 
from the far-infrared to the far-UV \citep[e.g.][ for a review]{draine03}. A sizeable fraction of the cold phase is locked up in dust grains
\citep[e.g.][ and references therein]{ss96,jenkins09}. Amorphous silicate materials together with graphite and policyclic aromatic
carbon should account for the majority of the depleted elements measured in the ISM: C, O, Fe, Mg, Si \citep[e.g.][]{wd01,wooden08}.
One of the major spectroscopic signatures of the presence of Fe- and Mg- rich amorphous silicates is the 10\,$\mu$m emission feature. 
Extensive studies of this feature 
lead to the conclusion that the Fe:Mg ratio should be approximatively 1
\citep[e.g.][]{lidraine01}. Main sources of both Mg and Fe silicates are O-rich asymptotic giant branch stars (AGB) and supernovae. However, the process of amorphization 
of the dust agglomerates, for instance by rapid cooling of the
gas phase \citep{wooden05} or cosmic rays bombardment \citep{carrez02}, strongly favors the survival of Mg silicates. 
In fact, a recent analysis successfully models the
 10\,$\mu$m feature in terms of Mg-rich silicates, if non-spherical grain shapes are used \citep{min07}. 
  Glassy material, consisting of Mg-rich silicates with metallic iron
 and sulfide inclusion \citep[called GEMS,][]{bradley94} have also been commonly found during the {\em Stardust} mission. Their origin is mostly from the interplanetary environment 
 \citep[e.g.][]{kel_mes04}, but a 
 fraction have a composition compatible with an ISM origin \citep{kel_mes08}. Iron is however a highly depleted element 
 \citep[70--99\% of Fe is in dust,][]{wilms00,whi03} whose inclusion into solid grains is not completely understood \citep[e.g.][]{whi03}. 
 This is mainly due to the difficulty of modeling iron emission, which does not display any sharp feature in long-wavelength spectra.
 
The abundances of the most important metals in the Galactic disk smoothly decrease with the galactocentric distance. The average slope of the distribution is 
$\sim0.06$\,dex\,kpc$^{-1}$ \citep[][ and references therein]{chen03}. However, a large scatter in the abundance measurements as a function of the Galactic radius 
is reported. This is attributed to different factors which contaminate the smooth mixing due to the pure stellar evolution process. Indeed the medium can be locally 
influenced by e.g. supernovae ejecta, and in-falling metal-poor material into the disk \citep{lugaro99,nittler05}. 

In recent years it has become clear that the X-ray band could provide an
excellent laboratory to study the silicate content of the diffuse ISM, as
the absorption K edges of O ($E=0.538$\,keV), Mg ($E=1.30$\,keV)
and Si ($E=1.84$\,keV) together with the Fe LII and LIII edges ($0.71$ and
$0.72$\,keV, respectively) fall in the low-energy X-ray band. The method used is to study the absorbed spectrum of bright X-ray binaries, located in different regions in the disk, 
observed with high-energy resolution instruments. This allows to probe the
interstellar dust (ID) content in a variety of environments, with different extinction and with possibly different dust formation history. 
Previous studies of X-ray spectra taken along different lines of sight led first to the recognition that not only gas but also dust plays a role in shaping the iron and oxygen edges 
\citep{takei02,kaastra09}
 and later led to the quantitative modeling of those edges \citep{lee09,devc09,ciro}. There is not yet a clear picture of the chemical composition of the ID as seen in X-rays. Silicates
containing andratite (iron-rich silicates) were reported studying the
oxygen edge of \object{GS~1826-238} \citep{ciro}, while iron oxides, rather than iron silicates were reported along the line of sight of \object{Cyg~X-1}
\citep{lee09}. This may point to a chemically inhomogeneous distribution of ID.\\

\object{4U~1820-30} is an extensively studied source, by virtue of its extraordinary intrinsic properties. 
It is an ultracompact \citep[orbital period 11.4\,minutes,][]{1987ApJ...312L..17S} X-ray binary 
consisting of a neutron star and a He white-dwarf \citep{1987ApJ...322..842R}. 
The presence of X-ray bursts associated with the neutron star has been early recognized \citep[first by][]{grindlay76}. \uu\ is classified as an atoll source \citep{1989A&A...225...79H},
showing kilohertz quasi-periodic oscillation at different frequencies in its power spectrum
\citep[e.g.][]{smale97,zhang98}.\\  
The broad band spectrum has been studied with several instruments. The source displays a Comptonized continuum and a soft black body component \citep[e.g.][]{sidoli01}.

With the advent of high resolution spectroscopy, the low-energy X-ray spectrum also revealed interesting features. 
Absorption by ionized ions has been reported in several studies \citep{futamoto04,yw05,juett06,cackett08}. The absence of blueshift in the
lines of this gas and lack of variability generally points to an interstellar origin \citep[but see][]{cackett08}. The hypothesis of a gas intrinsic to 
the source is intriguing as other X-ray binaries often display absorption by ionized gas either at the rest frame of the source \citep{vanpeet} or outflowing \citep[e.g.][]{jmiller06,neilsenlee09}. 
The source has
also been used as a backlight to illuminate the interstellar dust along the line of sight. This results in a dust scattering halo, which in this source is 
moderate \citep[$\sim3.2$\% at 1\,keV,][]{peter95}, given the relatively low 
Galactic column density ($N_{\rm H}\sim1.63\times10^{21}$\,cm$^{-2}$, this study).

Absorption by cold interstellar dust has been never studied in detail in this source. In this paper we aim for a comprehensive view of the cold absorbing medium along the line of sight 
to this source. This study benefits from the combined information provided by the \ch\ and \xmm\ high resolution instruments, which
allows us to meaningfully study both the Fe\,L and O\,K edges. In addition, we make use of dust models updated with the latest laboratory measurements 
\citep[e.g.][]{lee08,lee09}.
We also present
the broad band continuum behavior underlying the absorption features.\\

%__________________________________________________________________

The adopted protosolar abundances follow the prescription given by
\citet{lodders09} and discussed in \citet{lodders10}. The broad band energy spectrum (\epi\ and \inte) is fitted using the $\chi^2$ minimization method, taking care that at least 20 counts per
bin are present in each data set. For the high resolution spectra (RGS and \ch-HETG) the Cash statistic has been used\footnote{http://heasarc.gsfc.nasa.gov/docs/xanadu/xspec/manual/XSappendixCash.html}. The errors quoted
are for 68\% confidence level, corresponding to $\Delta\chi^2=1$ or $\Delta C^2=1$. The spectral fitting package used in this paper 
is SPEX\footnote{www.sron.nl/spex} \citep{kaastra96}. The adopted distance is 7.6\,kpc \citep{kuulkers03}. The nominal hydrogen column density toward the source is
$1.29\times10^{21}$\,cm$^{-2}$ \citep{kal05}, to be compared with $1.52\times10^{21}$\,cm$^{-2}$ \citep{dl90}. 

This paper is organized as follows: in Sect.~\ref{par:data}, we illustrate the data handling for the different instruments used. Sect.~\ref{par:cont} is devoted to the continuum
determination, using \epi\ and \inte\ data. In Sect.~\ref{par:high} we describe in detail the modeling of the different absorption components using RGS and \ch-MEG. The discussion can be found in
Sect.~\ref{par:discussion} and the conclusions in Sect.~\ref{par:conclusions}.\\

\section{The data handling}\label{par:data}
 
\subsection{Epic-pn}\label{par:epi}

\uu\ was observed by \xmm\ for 41\,ks on April 2, 2009 (revolution 1706). The data reduction was performed using SAS
(ver.~9.0). The \epi\ \citep{struder} was operated in full-frame masked
mode. In this mode, the central 13x13 (in RAW coordinates) pixels are flagged as bad on board, masking the central part of the PSF. The full-frame
mode was chosen to optimally study the halo of diffuse emission surrounding the source, caused by the scattering of the 
intervening dust in the line of sight. However, the spectrum of the central source is recovered from the so called
Out of Time (OoT) events, which are the photons trailed through the detector during the read out of the frame. The
exposure time of the OoT events is 6.3\% of the total exposure. We
extracted the source spectrum from the RAW coordinate event file. We selected 11 columns from RAWX 32 to RAWX 42, avoiding the central 2 columns, which were affected by pile
up. We selected RAWY$<$89 to avoid the innermost region of the PSF, 
which is still affected by pile up or X-ray loading\footnote{http://xmm2.esac.esa.int/docs/documents/CAL-TN-0050-1-0.ps.gz}.\\
The background was selected in a neighboring region. The exposure time of the background was scaled to match the
effective exposure of the OoT events. Given the brightness of the
source, the nominal background contribution is modest. However the whole
area is contaminated by the photons of the scattering halo, which have a
different distribution depending on the position in the detector. The OoT events were
extracted from a modified event file processed in such a way that the charge transfer inefficiency (CTI) corrections are not applied and all the
photons are considered as coming from the central source coordinates.

\subsection{RGS}\label{par:rgs}
 
\begin{figure}
\resizebox{\hsize}{!}{\includegraphics[angle=90]{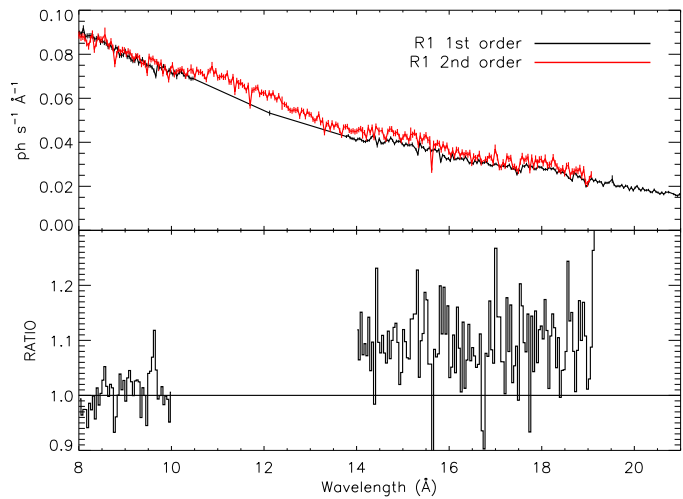}}
\caption{\label{f:rgs_ratio} Upper panel: comparison between the 2nd and 1st order for RGS1. Lower panel: the ratio between the 2nd and the 1st order provides an indication of pile up in
the spectrum.}
\end{figure}

Due to the brightness of the source, some RGS \citep{denherder} portions of the grating data were affected by pileup. 
In addition, in RGS2 the high
number of events per CCD frame (twice the exposure time compared to RGS1,
due to the single node readout) tended to fill up the buffers too fast.
Limits on the data handling in relation to telemetry prevented the buffers
to be cleared in time, leading to data losses for some CCD's. A method to
recognize the spectral regions not affected by pile up is to compare the first and the second order for each grating.
Calibration shows that in absence of pile up, the ratio of the spectra of the two orders, 
in the region where they overlap, should be unity\footnote{http://xmm.vilspa.esa.es/docs/documents/CAL-TN-0075-1-0.pdf}. In Fig.~\ref{f:rgs_ratio}, 
we show the ratio for RGS1. We see that the maximum pile up is recorded between 14 and 19 \AA,
where the effective area of RGS is the largest. In this region the ratio is about 1.1, which translate in only a 6--7\% pile
up. In RGS2 the ratio between 2nd and 1st order reaches even a value of five in selected regions. At longer wavelengths ($\lambda>20$\,\AA) the combined 
effect of absorption and decrease of the effective area lowers significantly the count rate. Therefore, this
region is not affected by pile up and can be safely used. Absorption
features are however in general minimally influenced by this effect. We have chosen to keep the useful data given by RGS2 around the iron L
edge, with care of locally fitting a different continuum for RGS1 and RGS2, using ``sectors" in SPEX, as defined below (Sect.~\ref{par:high}).\\
We also used archival RGS data (Table~\ref{t:log}) in order to improve the signal-to-noise (S/N) ratio. This observation
displayed a short period of flaring background, which was filtered out.
This resulted in a cut of 3\,ks on the total exposure time.  

\subsection{\ch-HETG}\label{par:ch}
In order to gain more signal to noise and energy resolution at wavelengths$\ltsim$18\AA, where the Fe, Mg and Si edges
stand, we considered archival \ch-HETG
data, selecting the higher-flux data sets available (Table~\ref{t:log}). We obtained the final products, processed in April 2010, from 
the TGCat archive$\footnote{http://tgcat.mit.edu/}$. The three observations displayed similar fluxes and continuum
parameters, therefore we combined the data, separately for the HEG and the MEG arms. HEG data are of low signal-to-noise at wavelengths
larger than $\sim$17\AA. In the following we refer to the MEG data only, unless otherwise specified.

\subsection{\inte}
In order to constrain the continuum spectrum of 4U 1820-30 across an as-large-as-possible energy band, INTEGRAL data 
can be very useful. Fortunately, the Galactic center region was observed by INTEGRAL for $\sim$ 73 ks from March 30, 2009 to March 31,
2009 during satellite revolution 789, and for $\sim$ 38 ks on April 3, 2009 during revolution 790. The XMM observation ended during 
the latter observation. Therefore we can consider these INTEGRAL observations as 
quasi-simultaneous with the XMM-observation.
4U 1820-30 was in the field of view of the INTEGRAL Soft Gamma-Ray Imager ISGRI \citep[15-300 keV;][]{lebrun03}
at source angles smaller than $14\fdg5$ for all 21 individual pointings (so called science windows) of the rev.~789 observation 
and all 10 pointings of rev-790.
The source was outside the field of view ($13\fdg2$ diameter zero response) of the Joint European Monitor for X-rays JEM-X 
\citep[3-35 keV;][]{lund03} during the rev.~789 observation, but within the field of view during 4 individual pointings of rev.~790.

We generated mosaic maps for 10 logarithmically spaced energy bands across the 20--300 keV range for the ISGRI rev.~789 observation 
using OSA version 9.0 \citep[distributed by the ISDC;][]{courvoisier03} imaging tools.
We checked these maps for significantly detected sources and subsequently derived spectral information for all these sources in 
13 pre-defined energy windows across 13--520.9 keV.

For JEM-X (telescope 1) we followed similar imaging and spectral extraction procedures yielding spectra of \uu\ in 16 pre-defined energy bands across the 
3.04--34.88 keV.The observation
log is displayed in Table~\ref{t:log}.

\begin{table}[t]
\caption{\label{t:log} The multi-instrument observation log for \uu.}
\begin{center}
\begin{tabular}{lccll}
\hline
\hline
Inst. & orbit/obsid & date  & net exp. & rate$^{1}$ \\
& &(dd/mm/yy) & (ks) & (c/s)\\
\hline
XMM-pn & 1706 & 02/04/09 & 39.8 & 20.9$^{2}$\\
\inte-JEMX & 790& 03/04/09 & 0.61& 28.8\\
\inte-ISGRI & 789& 31/03/09 &42.4 &15.5\\
\hline
XMM-RGS & 1706 & 02/04/09 & 41.5 & 27.6\\ 
XMM-RGS & 0336 & 09/10/01 &34.7 & 33.1\\
\ch-MEG & 6633 & 13/08/06 & 25 & 125.1\\
\ch-MEG & 6634 & 20/10/06 &25 & 172.8\\
\ch-MEG & 7032 & 05/11/06 &46 & 146.2\\
\hline
\end{tabular}
\end{center}
$^1$ Rates refers to the full band for each instrument.\\
$^2$ Rate from selected pixels of the OoT events.
\end{table}

\begin{figure}
\resizebox{\hsize}{!}{\includegraphics[angle=90]{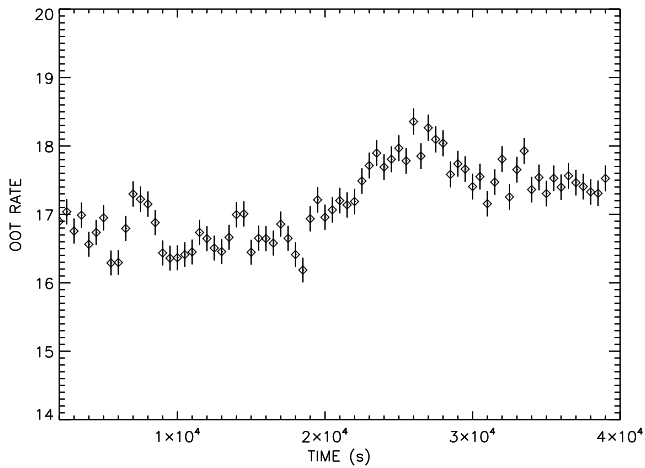}}
\caption{\label{f:lc} Light curve of \epi\ data as selected in Sect.~\ref{par:epi}. Note that the displayed count rate 
cannot be used to recover the flux of the source. The bin time is 500 s.}
\end{figure}
\begin{figure}
\begin{center}
\resizebox{\hsize}{!}{\includegraphics[angle=90]{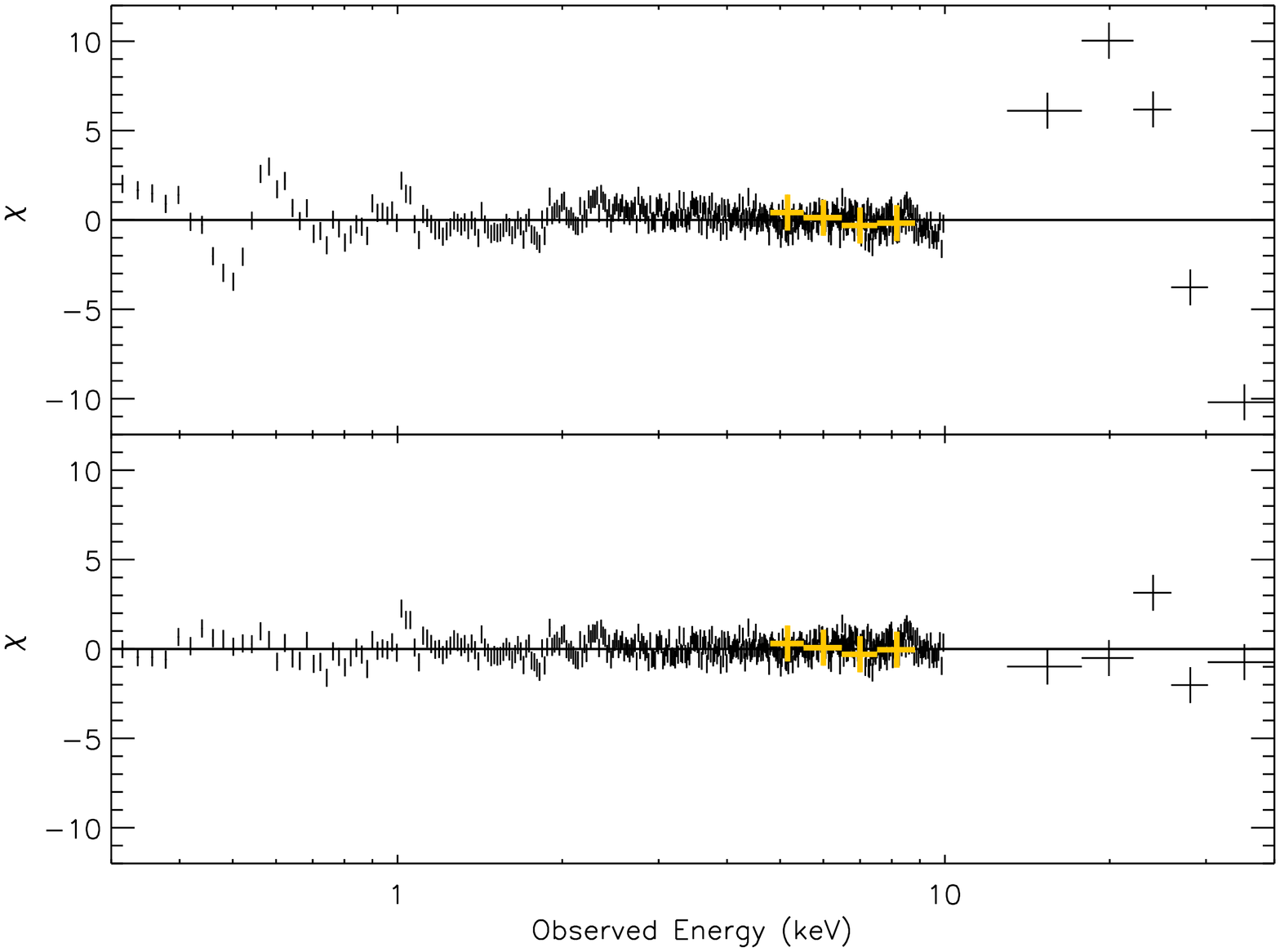}}
\end{center}
\caption{\label{f:broad_a} 
Upper panel: residuals to a simple absorbed power law for \epi\ and \inte\ data. JEMX data have been highlighted for
clarity. Lower panel: residuals to the best fit consisting of a black body plus Comptonized emission, modified by both a
neutral and ionized absorbing material. \epi\ data have been rebinned for displaying purpose.}

\begin{center}
\resizebox{\hsize}{!}{\includegraphics[angle=90]{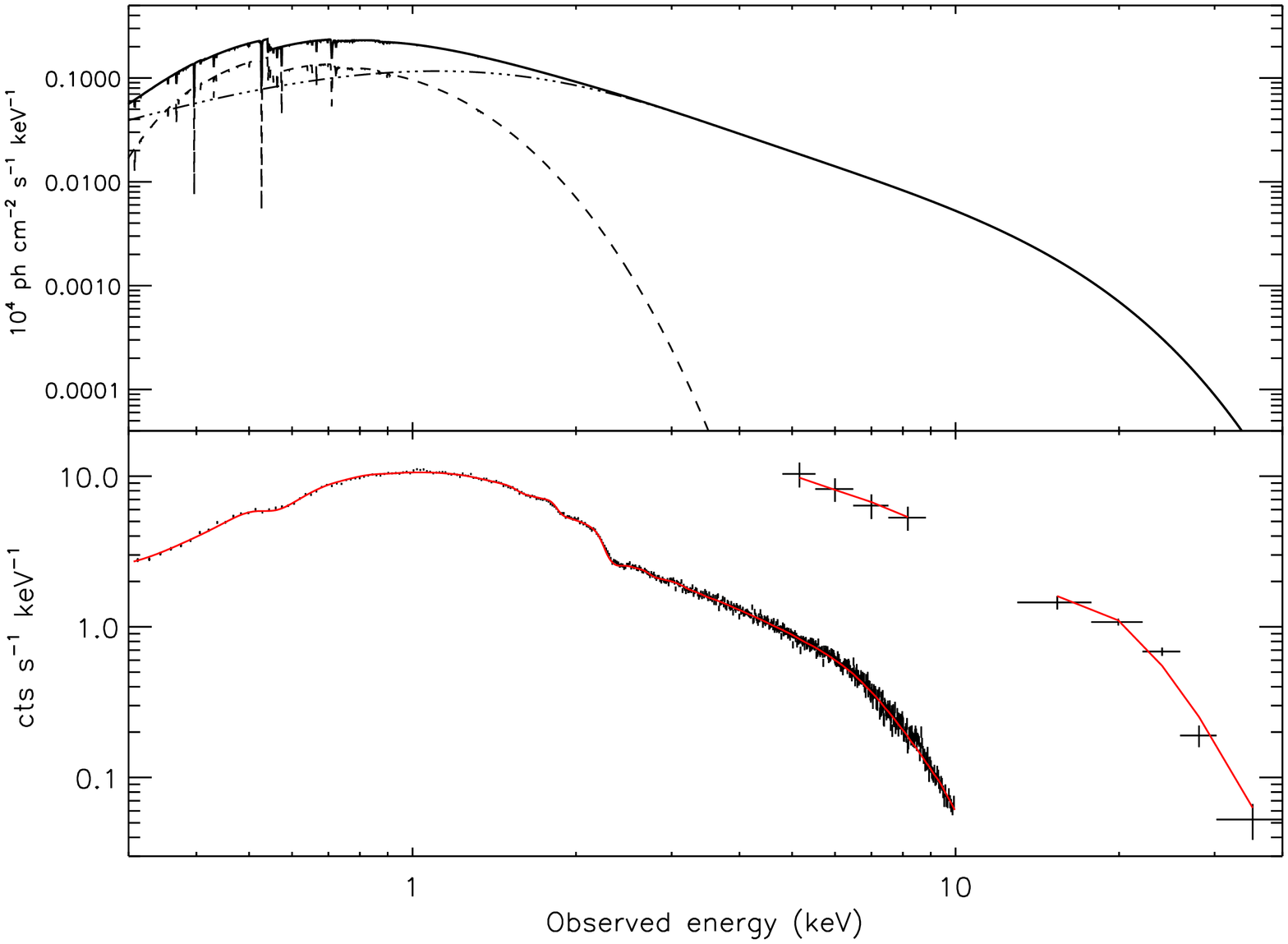}}
\end{center}
\caption{\label{f:broad_b} Upper panel: The best fit model consists of a black body (dashed line) 
plus Comptonized emission (dot-dashed line), modified by both a
neutral and ionized absorbing material (Table~\ref{t:broad_band}). Lower panel: Best fit model superimposed to 
the \epi,
 JEMX and ISGRI spectra.} 
\end{figure}

\begin{table}
\caption{\label{t:broad_band} Broad band modeling of the source using \epi\ and \inte\ data.}
\begin{center}
\begin{tabular}{ll}
\hline\hline
$N_{\rm H}^{\rm Gal}$  &  $7.8\pm0.3$\\
$T_{\rm bb}$ & $0.26\pm0.01$\\
$kT_0$ & $0.48\pm0.02$\\
$kT_1$ & $3.9\pm0.1$\\
$\tau$ & $5.4\pm0.2$\\
$\chi^2/\nu$ & 1756/1720\\
$F_{\rm 0.5-2\ keV}$$^1$ &$1.32\pm0.02\times10^{-9}$\\
$F_{\rm 2-10\ keV}$$^2$ & $5.02\pm0.07\times10^{-9}$\\
$F_{\rm 10-60\ keV}$$^3$ &$2.70\pm0.02\times10^{-9}$\\
\hline
\end{tabular}
\end{center}
Notes:\\
Units are: $10^{20}$ cm$^{-2}$ for column densities, keV for temperatures, Fluxes are measured in erg\,cm$^{-2}$\,s$^{-1}$.\\
$^1$ from RGS data\\
$^2$ from JEMX data\\
$^3$ from JEMX and ISGRI data.   

\end{table}

\section{The broad band spectrum}\label{par:cont}

We first extracted the light curve from the \epi\ OoT events as described in Sect.~\ref{par:epi}, 
therefore the resulting count rate cannot be used to recover the absolute flux. The 0.5--10\,keV light curve is shown in Fig.~\ref{f:lc}, 
using a bin size of 500\,s. It shows a modest variation of at most 13\% during our observation. The rates in 
the soft (0.5--2\,keV) and the hard (2--10\,keV) band followed the same pattern.\\  
We fitted the \epi\ and \inte\ spectra simultaneously. The source was caught at a relatively high flux, with a
2--10\,keV flux of $\sim 5.0\times10^{-9}$\,erg\,cm$^{-2}$\,s$^{-1}$ (Table~\ref{t:broad_band}). In Fig.~\ref{f:broad_a} (upper panel) 
we show the residuals with respect to a simple powerlaw with photoelectric absorption. 
From the residuals we see a complex spectrum at energies below 1\,keV, 
probably due to ionized absorbing
gas. The spectrum is well detected up to 40\,keV.\\
As a simple powerlaw is not an acceptable model ($\chi^2/\nu=2412/1724=1.39$, where $\nu$ 
is the number of degrees of freedom), 
we added first a black body component, to mimic the often observed soft energy
emission ($\chi^2/\nu=2273/1722=1.32$). This is parameterized by the black-body temperature ($T_{bb}$) and its normalization. 
However a black body plus power law model fails to explain the spectral curvature seen at high energies. 
We then substituted the powerlaw model with a Comptonization model 
\citep[COMT model in SPEX;][]{1994ApJ...434..570T}, where the parameters are: the temperature of the seed photons ($kT_0$), 
the optical depth of the electron cloud ($\tau$) where the photons are Compton scattered to the final temperature ($kT_1$). This model provides a satisfactory fit to the broad band continuum 
($\chi^2/\nu=1756/1720=1.02$; see Fig.~\ref{f:broad_a}, lower panel, Fig.~\ref{f:broad_b}, and Table~\ref{t:broad_band}). 

We repeated the fits using a disk black body model (i.e., emission from a standard Shakura-Sunyaev disk, model DBB in SPEX) instead of a simple black body.
We do not obtain an equally good fit using DBB either in addition to a simple powerlaw model ($\chi^2/\nu=2302/1722=1.33$) or DBB plus Comptonization 
($\chi^2/\nu=1998/1720=1.16$).\\

In the 6.4\,keV region, there is no clear evidence of iron emission lines. A 3\,$\sigma$ upper limit of 25\,eV on the equivalent width 
is obtained if a delta line is put at
a fixed energy of 6.4, 6.70 and 6.97\,keV, respectively. We also searched for a relativistically modified line profile \citep[LAOR model in SPEX; ][]{laor91}. 
In this model, the line arises from the accretion disk where the intensity follows a $R^{-q}$ profile, where $R$ is the distance of the disk gas from the source. 
Leaving the energy as a free
parameter did not lead to a meaningful fit. We then fix the energy of the line at
6.97\,keV, following previous studies of \uu\ \citep[e.g.][]{cackett10}. The best fit yelds a very broad, but weak line profile, with parameters $q=2.8\pm0.2$ and $i<14^{\circ}$, where $i$ is the
disk inclination. The flux of the line is $(1.6\pm0.7)\times10^{-4}$\,photons\,cm$^{-2}$\,s$^{-1}$.     

The complex absorption at soft energies has been modeled following the prescription given by the high-resolution data
(Sect~\ref{par:rgs}). Therefore we added a minor contribution from ionized absorbers ($N_{\rm H}^{\rm ion}\sim {\rm few}\times10^{20}$\,cm$^{-2}$) and the Galactic neutral absorber, which
contributes to most of the low-energy curvature. We note that the best fit neutral column density is
about 40\% lower than the best fit obtained in the RGS analysis. This may be caused by foreground scattering halo soft-emission (observed for the full exposure) 
located just in front of the OoT event (observed for 6.3\% of the exposure time) which 
still remains after the subtraction of the local background. The scattering halo appears indeed as diffuse emission which
extends on top of both the wings of the point spread function of the source and the OoT events themselves 
\citep[e.g.][]{peter95,costantini05}. The scattering process is strongly energy dependent \citep{mathis_lee91}. In
particular, the halo is brighter at softer energies. The net effect is therefore to add more photons to the soft energy
spectrum of the OoT events, reducing the measured absorption toward the source. This effect appears even more enhanced by the
reduction of the OoT absolute flux due to the cut of the central two columns, where most of the source photons are.

\section{The high resolution spectrum}\label{par:high}
Since the energy band of the RGS is insufficient to discriminate among different broad-band models, 
we adopted the model defined in the previous section. The spectrum shows evident \oi\ K and \fei\ L edges, due to absorption by neutral material. In addition, absorption by 
ionized gas is highlighted by the \oviii, \ovii\ and \ovi\ absorption lines.\\
We used a collisionally ionized plasma model (model HOT in SPEX), with a temperature frozen to $kT=5\times10^{-4}$\,keV, in order
to mimic a neutral gas. This model fits to first order the edges and resonant lines from the neutral species, leaving
however noticeable residuals in the fit. 
In order to extract more robust results from the absorbtion features, we added archival RGS (Sect.~\ref{par:rgs}) and \ch-MEG data (Sect.~\ref{par:ch}). 
As these data sets were obtained at
different epochs, the continuum shape may differ significantly. To bypass this complication, we used the "sectors" option is
SPEX\footnote{See Ch.~5 of the SPEX cookbook: http://www.sron.nl/files/HEA/SPEX/manuals/spex-cookbook.pdf} and we left the
continuum parameters free to vary for each sector. Note that the RGS1 and RGS2 of the recent observations were
treated as different data sets (thus with different continua), as the RGS2 broad band shape was affected by pileup.    
The parameters of absorber components were coupled together for all the data sets.

\subsection{The ionized gas}\label{par:ovii}
The RGS spectrum of \uu\ displays narrow absorption features from ionized gas, 
mainly from oxygen, iron and neon. In particular \neix, \ovi --\oviii\ are prominent features of this gas. 
For this fit we ignored the data below 13\,\AA\ for RGS. In this region some wiggles in the local continuum, mainly caused by pile up,
could add uncertainties in the determination of the absorption parameters. 
Here we focus on the neutral or mildly ionized phase of the ISM. Therefore we simply model the higher ionization 
lines with a phenomenological model (SLAB in SPEX) which 
calculates the transmission from a thin layer of gas. We defer a more detailed analysis of this gas component to a future publication. 
The resulting ionic column densities of the main ions, as measured by RGS, are reported
in Table~\ref{t:ovii_slab}. Following \citet{yw05}, we assume a velocity broadening of $\sigma_v=62$\,km\,s$^{-1}$. All subsequent fits in the present analysis 
already take into account this highly ionized component. 

\begin{table} 
\caption{\label{t:ovii_slab} Parameters of the main lines of the more ionized ions as measured by RGS.}
\begin{center}
\begin{tabular}{ll}
\hline\hline
ion & log$N_{\rm ion}$\\
& cm$^{-2}$\\
\hline
\cvi & $16.7\pm0.1$\\
\nvi & $<15.1$\\
\nvii & $15.5\pm0.4$\\
\ovi & $15.4\pm0.3$\\
\ovii &$16.2\pm0.1$\\
\oviii & $17.6\pm0.1$\\
\hline
\end{tabular}
\end{center}
Note:\\ 
The assumed velocity broadening is $\sigma_v=62$\,km\,s$^{-1}$.\\

\end{table}

\subsection{The oxygen edge}\label{par:oxy}

\begin{figure}
\begin{center}
\resizebox{\hsize}{!}{\includegraphics[angle=90]{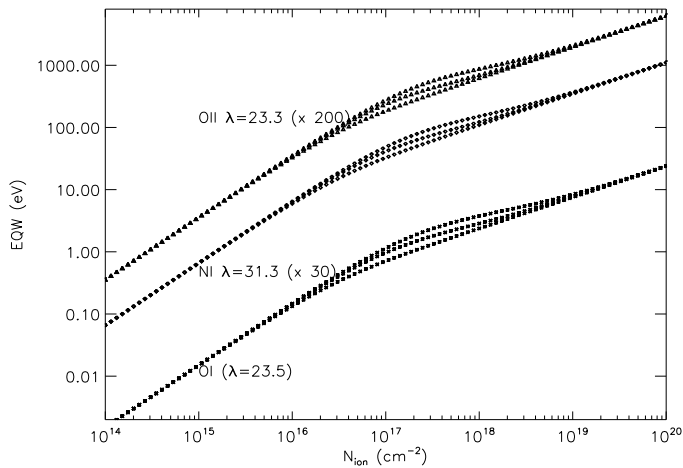}}
\end{center}
\caption{\label{f:voigt} Line equivalent width as a function of the ionic column density for \ion{O}{i}, \ion{N}{i} and \ion{O}{ii}. The velocity dispersions are $\sigma_v=$10, 100,
200\,\kms. Curves of different ions have been shifted for clarity.}
\end{figure}

\begin{figure}
\begin{center}
\resizebox{\hsize}{!}{\includegraphics[angle=90]{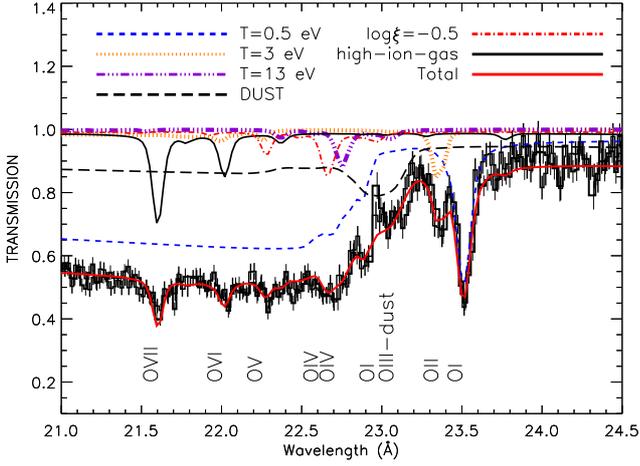}}
\end{center}
\caption{\label{f:oxy_best} Best fit transmission spectrum (Table~\ref{t:oxy}, column {\it (3)}) in the oxygen edge region. This allows to compare absorption at different epochs, removing the
contribution of the continuum. The upper curves display the transmission of the various
absorbing components. Note that for clarity only absorption by oxygen is displayed (i.e. no hydrogen absorption). Data have been rebinned for clarity.}
\end{figure}

The oxygen edge at 0.538\,keV has been fitted using the high quality data of the two RGS epochs between 19--36\,\AA. 
The neutral material, modeled by a low temperature 
($kT=5\times10^{-4}$\,keV) in a collisional ionization
equilibrium, fits well the long wavelength curvature due to the ISM absorption as well as the \oi\ 1s-2p absorption line at 23.5\,\AA\  (Fig.~\ref{f:oxy_best}). The edge shape
appears modified by several sharp absorption features. Some of them can be easily identified with the \ovii\ and \ovi\
lines, belonging to a highly ionized phase (Sect.~\ref{par:ovii}). At the shorter wavelength side of the \oi\ line, we see \oii\  at
$\lambda=23.35$\,\AA\ and also weaker \oiii\ and \oiv\ lines. Two absorbers with different temperatures are required to fit these lines (Table~\ref{t:oxy}, column {\it (1)}). 
The reported
line widths are automatically evaluated in SPEX using a curve of growth analysis applied to multiple lines belonging to the same absorption system \citep[e.g.][]{spitzer}. 
This allows to accurately evaluate the line width using basic parameters such as the ionic column density and equivalent width. In this case, \ion{O}{i}, \ion{Fe}{i}, \ion{N}{i}
transitions are used for the cold gas (comp~1 in Table~\ref{t:oxy}). The strongest transitions are saturated, which in principle may introduce additional uncertainty in determining the
line width. However, as shown in Fig.~\ref{f:voigt}, for the lines we study here, the ionic column density as a function of the line equivalent width depends 
marginally on the velocity width. This behaviour of the curve of growth is due to the large $a$-Voigt parameter for 
the inner-shell transitions \citep[][ for a full discussion]{mihalas,kaastra08}. 
We note that the nitrogen region is not modified by dust and the line widths are then easier to evaluate. In addition, many of the the weaker transitions 
of the same neutral ions are not saturated.  For the $kT\sim3.2$\,eV gas (comp~2), \ion{O}{ii}, \ion{Fe}{ii}, \ion{N}{ii} are among the strongest ions. These are relatively 
weaker lines where saturation does not play a major role. The relatively limited resolution of the instrument does not
allow to distinguish multiple velocity-width components within a same absorption line. Therefore we consider here the total ionic column density along this line of sight.\\

Dust should also be present along this line of sight, as \uu\ is known to display a dust scattering
halo \citep{peter95}, which is also evident in the present \epi\ data. 
We therefore fitted the oxygen edge adding the AMOL model in SPEX. We
considered 16 oxygen compounds for this analysis (see Appendix A). This list includes silicates and oxides which are common in
the ISM plus some lighter (also icy) materials. Absorption features 
by dust, contrary to gas, are smooth and broadened. All the oxygen compounds considered here have no appreciable effect
at $\lambda\ltsim 23.7$\,\AA\ (Fig.~\ref{f:appendix}).\\
Below we describe a simple fit using a single dust component at a time, adding a second component if the fit requires it. This makes the description of the model and of the steps to
the final best fit simpler to follow. This approach is however the result of a rigorous fitting procedure for both the oxygen and iron edges, 
which is fully described below (Sect.~\ref{par:fe_oxy}).\\

\begin{figure}
\begin{center}
\resizebox{\hsize}{!}{\includegraphics[angle=90]{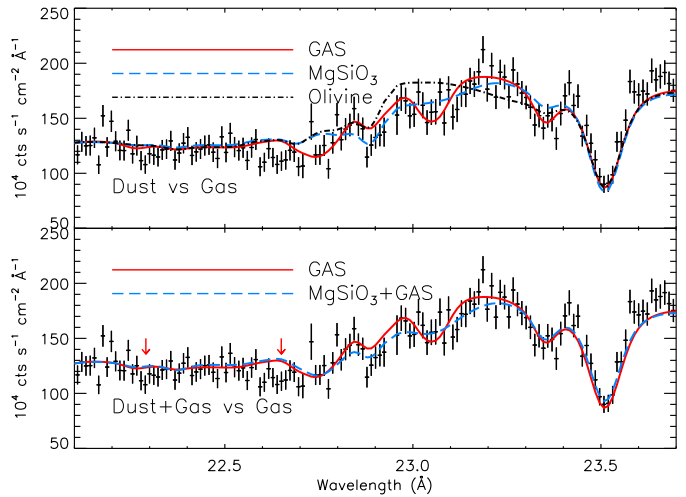}}
\end{center}
\caption{\label{f:oxy_bc} Fits in the oxygen region. Here only RGS1 of the most recent observation is displayed for clarity. Upper
panel: comparison between a fit only in terms of gas (solid line) and the best fit obtained using dust (dashed line). 
For comparison
also an unacceptable fit using olivine (dashed-dotted line) is displayed. 
Lower panel: comparison between a fit only in terms of gas (solid line) and the best fit obtained with a  
mixture of gas and dust (dashed line). See more details in the text.}
\end{figure}

We first attempted to fit the whole spectral region only in terms of dust absorption. In
the fit we let the hydrogen column densities of the gas producing \oi\ and \oii\
(comp 1 and 2 in Table~\ref{t:oxy}) as free parameters.
We also let the abundance of oxygen in the cold phase (comp 1) as a free parameter, in order to balance the amount of
oxygen locked up in dust. In this model, we suppressed the gas component (comp 3) mainly producing \oiii\ ($\lambda=23.07$\,\AA), as this line is the main absorption structure in the
region where also dust plays a major role.  
It is clear that fitting the region around 23\,\AA\ only in terms of dust 
remarkably worsens the fit. The worsening range is indeed $\Delta C^2=35-180$, depending on the compound, for $\Delta\nu=2$, 
with respect to the pure gas modeling. This indicates that not only dust is absorbing the spectrum in the $\sim22.7-23.1$\,\AA\ region. 
In Fig.~\ref{f:oxy_bc}, upper panel, we show the oxygen region where the pure gas fit is compared with the fit with MgSiO$_3$ (dashed line) 
or Mg$_{1.6}$Fe$_{0.4}$SiO$_{4}$ (olivine, dashed-dotted line), suppressing comp.~3. 

Finally, 
we included both the $kT\sim13$\,eV gas absorption (comp 3) and dust absorption (Table.~\ref{t:oxy}, column {\it (2)}). The improvement of the fit is 
then $\Delta C^2=10-108$ for $\Delta\nu=1$. Most of the individual dust components
improve the fit. However, we can identify two compounds for which the improvement of the fit is the largest 
($\Delta C^2=105-108$). These are H$_2$O, in the form of crystal or amorphous ice, and MgSiO$_3$. 
In Fig.~\ref{f:oxy_bc} (lower panel) we show how a fit with pure gas and a gas+MgSiO$_3$ mixture (comp 5 in Table~\ref{t:oxy}) compare with each other.
 Note that a fit using H$_2$O is non distinguishable in practice from MgSiO$_3$ as 
the spectral feature is very similar (Fig.~\ref{f:appendix}, Sect.~\ref{par:fe_oxy} for discussion). 
Interestingly we note that the compounds with both oxygen and 
iron provided the least improvement (if not a worsening) of the fit. In particular, the fit rules out the most 
complex aggregates of oxygen and iron (e.g. magnetite, franklinite, olivine and almandine, see Fig.~\ref{f:appendix} 
for the chemical composition).\\

As noted above, none of the dust compounds considered here have features below 23.7\,\AA. However, from 
Fig.~\ref{f:oxy_bc} we note two clear absorption features (present in
both RGS data sets) at $\sim22.3$ and $22.65$\AA\ (marked with arrows) detected with 6.8 and 3.8$\sigma$ significance, respectively. 
These features 
do not belong to the ionized absorber producing the \ovii\ line (Sect.~\ref{par:ovii}), but to 
a lower ionization gas. We tentatively fitted these features with a photoionization model (XABS in XSPEC), obtaining a low column density ($N_{\rm
H}\sim4.3\times10^{19}$\,cm$^{-2}$) and a ionization parameter\footnote{The ionization parameter $\xi$ is defined as $\xi=L/nr^2$, where L is the ionizing
luminosity, $n$ the gas density and $r$ the distance of the gas from the source.} log$\xi\sim-0.5$ (comp 6 in Table~\ref{t:oxy}). 
Interestingly, the two major lines, identified as
\oiv\ and \ov\ present a
systematic blue-shift of about $-1200$\,\kms\ (Table~\ref{t:oxy}, column {\it (3)}). From
the absorber model we notice that the other lines predicted using these
parameters fall either in a lower-resolution and noisier part of the
spectrum (e.g. lines of iron) or where the spectrum is heavily absorbed by
the neutral gas (e.g. nitrogen lines).\\ 
The best fit in the oxygen edge region, including RGS
data sets taken in two epochs is displayed in Fig.~\ref{f:oxy_best}.

\begin{table}
\caption{\label{t:oxy}Parameters for the oxygen edge modeling.}
\begin{center}
\begin{tabular}{lllll}
\hline\hline
&& {\it (1)} & {\it (2)}& {\it (3)}\\
\hline
{\bf comp 1} & $N_{\rm H}$ & $19.6\pm0.2$ & $17.3\pm0.2$ & $16.3\pm0.2$\\
& $T^{cold}$ & 0.5 fix. & 0.5 fix.& 0.5 fix.\\
& $\sigma_v$ & $20\pm10$ & 20 fix. & 20 fix.\\
{\bf comp 2} & $N_{\rm H}$ & $0.50\pm0.09$ & $0.56\pm0.09$ & $0.61\pm0.09$\\ 
& $\sigma_v$ & $<10$ & 10 fix.& 10 fix.\\
& $T$ & $3.2\pm0.4$ & $3.2\pm0.4$& $3.3\pm0.4$\\
{\bf comp 3} & $N_{\rm H}$ & $0.90\pm0.08$ & $0.36\pm0.06$ & $0.26\pm0.06$\\
& $T$ & $9\pm2$ & $13\pm3$ & $13\pm3$\\
& $\sigma_v$ & 100 fix. & 100 fix. & 100 fix.\\
{\bf comp 5} & $N_{\rm O}$ & $\cdots$ & $2.2\pm0.2\times10^{-3}$ &$2.3\pm0.2\times10^{-3}$ \\
 & $N_{\rm Si}$ & $\cdots$ & $7.4\pm0.8\times10^{-4}$ & $7.8\pm0.8\times10^{-4}$\\
 & $N_{\rm Mg}$ & $\cdots$ & $7.4\pm0.8\times10^{-4}$ & $7.8\pm0.8\times10^{-4}$\\
{\bf comp 6} & $N_{\rm H}$ & $\cdots$ &$\cdots$ & $0.43\pm0.06$\\
& log$\xi$ & $\cdots$ &$\cdots$ & $-0.5\pm0.$1\\
& $v_{\rm out}$ &$\cdots$  &$\cdots$ & $-1200\pm130$\\
& $C^2/\nu$ & 5863/4572 & 5745/4571 & 5684/4568\\
\hline

\end{tabular}
\end{center}
Notes:\\ 
Units are $10^{20}$\,cm$^{-2}$ for column densities ($N_{\rm H}$), eV for temperatures ($T$) and \kms\ for line
broadening ($\sigma_v$) and the outflow velocity ($v_{\rm out}$).\\
Column{\it (1)}: model with neutral and mildly ionized gas. Column {\it (2)}: mildly ionized gas + dust. Column {\it (3)}: milidly ionized gas + dust + mildly ionized
outflowing gas. In all models the highly ionized component is already included.  
\end{table}

\subsection{The iron L edge}\label{par:iron}
\begin{figure}
\begin{center}
\resizebox{\hsize}{!}{\includegraphics[angle=90]{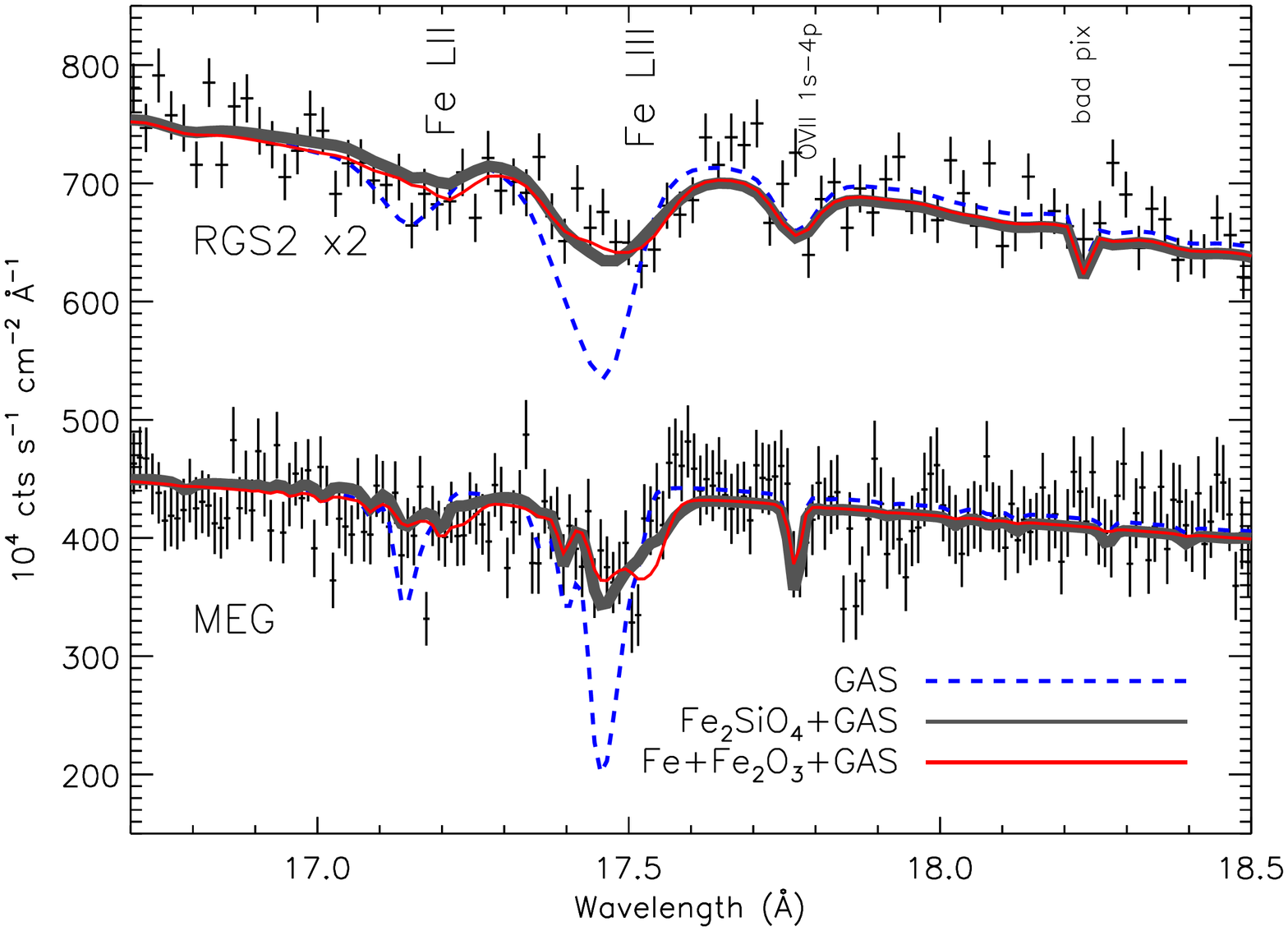}}
\end{center}
\caption{\label{f:fe_hot} Detail of the iron L edge region. Here for clarity we display only the MEG and a displaced RGS2
data set. The best fit (solid line, see Table~\ref{t:fe_fit}) is compared with pure gas fit with solar abundances
(dotted line) and with a mixture of gas and iron-rich olivine (thick solid light line).}
\begin{center}
\resizebox{\hsize}{!}{\includegraphics[angle=90]{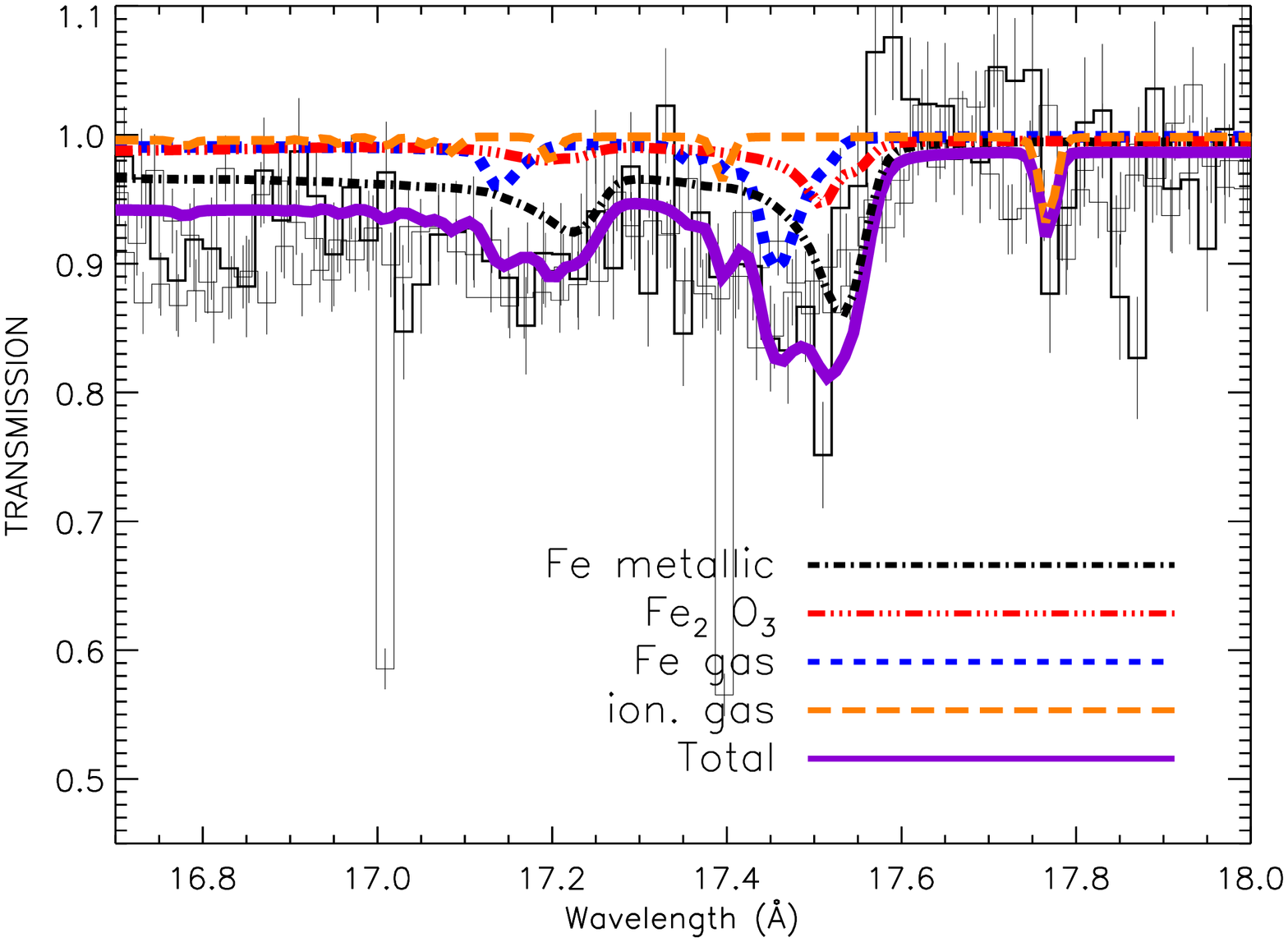}}
\end{center}
\caption{\label{f:fe_best} Best fit transmission spectrum of the region around the iron LII and LIII edges. This allows to compare absorption at different
epochs, removing the contribution of the continuum. The upper curves display the transmission of the various
absorbing components.}
\end{figure}

In Fig.~\ref{f:fe_hot}, a fit of the Fe LII and LIII edges in terms of absorption by pure gas with solar abundances 
is shown (dotted line). In this fit two RGS epochs and the MEG data sets were used. There is a clear mismatch not only in the depth, but also in the
position of the of the LII and LIII edges. We note that in the fit of the iron edges, it is essential to include the
higher resolution MEG data (Fig.~\ref{f:fe_hot}). 
A straightforward way to improve the fit is to modify the iron abundance. Letting this parameter free, we obtain a
reasonable fit, although the position of the edge in the model still does not match the data. In this case the abundance
of iron, assuming that the absorption is only in the gas phase, is about 0.37 times the solar one. 
However, absorption
by dust is known to alter the LII/LIII ratio and to shift the position of the edge with respect to absorption by gas
\citep[see e.g.][]{lee09}. For the fitting, we considered the compounds measured by
\citet{lee09}: metallic Fe, 
hematite (Fe$_2$O$_3$), lepidocrocite (FeO(OH)), fayalite (Fe$_2$SiO$_4$) and iron sulfate (FeSO$_4$). These 
transmission models have been recently implemented in the AMOL model in SPEX. As for the modeling of the oxygen edge, we first followed a rigorous approach (Sect.~\ref{par:fe_oxy}) which
in turns justifies the simpler approach described here, i.e. fitting the dust
components one by one, together with the gas model. In the latter, the iron abundance is a free parameter. This procedure 
provides a measure of the depletion of the gas phase.  On the goodness of fit bases, the single dust compound which best models
the data is metallic iron (Table~\ref{t:fe_fit}). Then we tried all possible combinations of gas plus two dust
components. We only find a marginally significant presence of Fe$_2$O$_3$, in addition to the metallic 
iron component (Table~\ref{t:fe_fit}, Fig.~\ref{f:fe_hot}, solid line). Hematite (Fe$_2$O$_3$) is a compound for which both iron and oxygen edges'
dust profiles are implemented in the AMOL model. We therefore tested the presence of this compound in the oxygen edge region,
obtaining an upper limit on the column density, consistent with the value obtained in the iron region. In 
Fig.~\ref{f:fe_hot} we also show the comparison between the best fit model and a model with an iron rich olivine 
(fayalite, \fay), which does not provide a good fit (Table~\ref{t:fe_fit}). This suggests that fayalite cannot have a
major contribution to the absorption here. We cannot test the same kind of olivine as for the oxygen edge
(Mg$_{1.6}$Fe$_{0.4}$SiO$_{4}$) as at present no laboratory measurement of this compound at the iron L edge is 
available. In Fig.~\ref{f:fe_best}, the best fit is shown where data from both RGS and
\ch-MEG are displayed. The spectra have been normalized to their continuum shape for displaying purposes. We also
normalized to the total hydrogen column density so that only absorption by iron is visible for the neutral phase. Along
with the best fit (solid line), also the different contribution are shown: atomic gas (dotted line) and metallic iron and 
Fe$_2$O$_3$ (light dash-dotted lines). Also the contribution of the ionized gas to the spectrum is shown with a dashed
line (e.g. \ovii\ line at 17.76\,\AA).\\ 
Despite the significant improvement of the fit with respect to a gas-only model, the iron edge is clearly more complex
that our parameterization. In particular, positive residuals are present on the longer wavelength side of the edge. This
effect has been found previously using different instruments \citep{kaastra09,lee09}.

\begin{table}
\caption{\label{t:fe_fit}Goodness of fit for the iron edge models.}
\begin{center}
\begin{tabular}{ll}
\hline\hline
model & $C^2/\nu$\\
Gas ($A_{\rm Fe}\ {\rm fix.}$ ) & 2377/981\\
Gas ($A_{\rm Fe}\ {\rm free}$) & 1544/980\\
\fay & 1505/977\\
\irsul & 1486/977\\
\lep & 1464/977\\
\hem & 1442/977\\
Fe met. & 1424/977\\
Fe met.+\hem & 1421/976\\
\hline
\end{tabular}
\end{center}
\end{table}

\subsection{Further insight in the Fe\,L and O\,K edge fitting}\label{par:fe_oxy}
The modeling of photoelectric edges modified by dust absorption requires some caution, 
as in principle many components can contribute to the edge shape, which is in turn smeared, contrary to the atomic
transitions features, which are very sharp and relatively easy to identify. Therefore the accuracy of the result depends on the resolution of the instruments, the signal-to-noise ratio, the completeness of the
dust models and finally on how many independent parameters there are in the problem.\\
In this study we combined the higher resolution of \ch-MEG with the sensitivity of RGS to study the Fe\,L edges. For the oxygen edge we used the high quality RGS data, as the \ch-HEG effective
area dramatically drops in that spectral region. \uu\ is also a high flux source, absorbed by a gas and dust column density which is optimal in order to study the O and Fe edges.\\
The dust edge profiles that we use in our study are a collection of what is available in the literature up to now \citep[e.g.][]{barrus79,parent02,vaken,lee08,lee09}. This data-base 
is not complete, but the main representatives of oxides and silicates for both oxygen and iron are present \citep[see Appendix~A, ][]{lee09}. This is sufficient for a first
order analysis.\\
The simultaneous analysis of two edges produced by dust on the same line of sight allows to cross-validate our results. In addition, common knowledge on abundances and physical conditions
in the ISM allow us to put further constraints to the edges fitting. Here we illustrate this in detail.\\

In the ISM a mixture of chemical compounds exists. We tried to simulate this by mixing all possible combinations of edges profiles available for a given edge. The AMOL model allows to fit
four different components at a time. We therefore had for each edge a number of fits $n$ where $n=nd!/4!(nd-4)!$. Here $nd$ is the number of available edge's profiles. Examining 
the output we notice that the combinations which improve the fits prefer one or two components, putting the remaining column densities to zero. If two components are preferred, one
strongly dominates over the other, whose relative value is never more than 20\%. This applies to both the oxygen and the iron edge. This justifies the simplified way of illustrating the
fitting used in Sect~\ref{par:oxy}, \ref{par:iron} above.\\

We further analyzed all fits that were within 3\,sigma the best fit values shown in Sects.~\ref{par:oxy} and \ref{par:iron}. Interestingly, for both iron and oxygen, 
no models containing iron-rich silicates
are among these selected models.\\
For iron we could rule out models, within the ``3\,sigma" set, containing hematite (Fe$_2$O$_3$) with a total molecular column density\footnote{Note that this translates in
$N<1.3\times10^{16}$\,cm$^{-2}$ and $N<8.8\times10^{15}$\,cm$^{-2}$ for O and Fe respectively.} larger than $4.4\times10^{15}$\,cm$^{-2}$. This resulted from the cross-validation we performed
with the oxygen edge, as for hematite we have both the O and Fe dust profiles. Indeed, none of the ``3\,sigma" models for oxygen contained this compound. 
We therefore added it to the oxygen edge best
fit model, obtaining the limit above (and compatible with what found for the iron edge best fit, see Sect.~\ref{par:iron}).\\ We could also rule out the "3\,sigma" models containing 
Fe$_2$S$_2$ as for this case the fit delivered an amount of sulfur $\sim$3.8 times solar. The sulfur edge would have been significantly enhanced, but this is
not evident in our data.\\
For the oxygen edge this multicomponent fitting pointed very clearly to a predominance of MgSiO$_3$ in all the "3\,sigma" set. As noted in Sect.~\ref{par:oxy}, a similarly good fit is
obtained if water ice is the predominant component. This is because of the very similar edge profiles of the two compounds (Fig.~\ref{f:appendix}). 
However, we could readily rule out this option,
as along the line of sight of \uu\ the bulk of cold material is diffuse interstellar medium, where water is virtually absent and silicates are on the contrary abundant
\citep[e.g.][]{wooden08}.

\subsection{The magnesium and silicon edges}\label{par:si}
The Mg K and Si K edge (at 9.47 and 6.723\,\AA, respectively) were analyzed using the combined information of \ch-HEG and
MEG, which provide in this band a higher energy 
resolution. In the case of Si, the energy of the edge is also outside the RGS range.
The quality of the data and the relatively low Galactic column density hampered any detailed characterization of the
edges. We used the auxiliary information provided by the oxygen edge fitting (Sect.~\ref{par:oxy}) where 
a model including MgSiO$_3$ was preferred. For that compound we
could benefit of the laboratory measurements of both the O and Si edge.
For Mg, for which we did not have any measurement, the relative column
density was included in the model in the form of an artificial
gas-shaped edge. We therefore added to the gas absorption model
this specific dust component, fixing the amount of magnesium and silicon locked in dust and letting the 
amount of Mg and Si in the gas phase
as free parameters. This model agrees with the data, within the uncertainties. The column density of Si in dust is 
$N_{\rm Si}^{\rm dust}\sim7.8\times10^{16}$\,cm$^{-2}$, while the Si in gas is  
$N_{\rm Si}^{\rm gas}<1.3\times10^{16}$\,cm$^{-2}$. Therefore the inclusion of Si in dust is 
$\sim$85\% of the total neutral Si along the line of sight. 
For the Mg edge, we obtain a  $N_{\rm Mg}^{\rm gas}<3\times10^{15}$\,cm$^{-2}$, while the value imposed by the oxygen fit to
the Mg locked in dust is 
$N_{\rm Mg}^{\rm dust}=(7.8\pm0.8)\times10^{16}$\,cm$^{-2}$. 
The amount of Mg locked up in dust according to this model is about 96\%.

\subsection{The neon and nitrogen edges}\label{par:nitrogen}
In order to study the neon edge ($\lambda=14.24$\,\AA), we used the combined data of RGS and \ch, while for the nitrogen edge we used only the
RGS data. These elements are only present in gas form in the ISM \citep[e.g.][]{wilms00}. 
Fitting neon with our Galactic absorption model we find a column density of $2.07\pm0.02\times10^{17}$\,cm$^{-2}$, which converts in 
a slight formal overabundance with respect to solar $1.2\pm0.1$.\\ 
Next to the nitrogen edge at 30.76\,\AA, we can also distinguish absorption by the 1s-2p 
line at 31.28\,\AA. The measured column density of nitrogen is
$(1.33\pm0.01)\times10^{17}$\,cm$^{-2}$, corresponding to an abundance of $0.9\pm0.1$ times solar, consistent with the value expected using solar abundances.

\section{Discussion}\label{par:discussion}

\subsection{The broad band continuum and the iron emission line}
We have analyzed the simultaneous observation of \uu\ using XMM-\epi\ and \inte. The fit of the continuum is consistent
with black body emission plus a Comptonized component, the latter extending up to 40\,keV. The source was observed in a
high flux state, accreting at $L_{\rm bol}/L_{\rm Edd}\sim0.16$. We have assumed $L_{\rm bol}=2\times L_{(2-20\ {\rm keV})}$, where the unabsorbed luminosity $L_{(2-20\ {\rm keV})}$ is 
$\sim3.1\times10^{37}$\,erg\,s$^{-1}$. The Eddington luminosity for ultracompact systems is $L_{\rm Edd}\sim3.8\times10^{38}$\,erg\,s$^{-1}$, following \citet{kuulkers03}. 
The spectral shape is also consistent with a high state
scenario. The hard energy portion of the spectrum is well fitted by a Comptonization model with an effective cut-off at 
$\sim40$\,keV, consistent with previous observations of the source in the high-state \citep{bloser00,sidoli01}. The
black body emission, which fits well the soft energy range of the present observation, has a temperature 
similar to that of the comptonization seed photons. This may suggest that in ultra-compact X-ray binaries 
seed photons originate from the accretion disk \citep{sidoli01}.

A relatively weak or undetected iron line is a common feature of the confirmed ultra-compact X-ray binaries
\citep[as defined e.g. in][]{zand07}, as \uu\ is. This may be simply explained by the nature of the donor star in such systems. Due to 
the high surface gravity (log\,$g\sim8$), heavier elements in the white-dwarf companion sink in the internal layer of 
the star \citep[e.g.][]{fm79}, leaving little high $Z$ metals for the accretion flow. 
In \uu\ however, a faint, variable and relativistically smeared iron line as been reported 
\citep[e.g.][]{cackett08b,cackett10}. In our data, no narrow emission features, either from neutral or ionized iron, have been found (Sect.~\ref{par:cont}).  
A fit with a relativistically smeared iron line leads to a very elusive 2-sigma detection of the feature.  
The photon flux of this line is about 16 times fainter than that found in recent $Suzaku$ data  \citep[][]{cackett10}.\\ 
In exceptional circumstances, like superburst episodes, a large quantity of heavy elements ashes may be ejected 
during the neutron star photosphere expansion. This could possibly cause detectable features (e.g. in absorption) by e.g. iron in the X-ray spectra
\citep{jean10}. In \uu\ the presence of the iron line could then be the tracer of burst activity, not detectable at the epoch of our observation.

\subsection{The dust components}
The high resolution spectrum of \uu\ shows evidence of absorption by dust, especially around the Fe\,L and O\,K edges
(Figs.~\ref{f:oxy_best}, \ref{f:fe_best}). In the fits we explored all the possible combinations of dust mixtures (Sects.~\ref{par:iron} and \ref{par:oxy}). 
The best fit shows a preference for absorption by enstatite (\mg), which is the end series of the pyroxene
silicate mineral, metallic iron (Fe) and traces of iron oxides in the form of hematite (\hem). In the modeling, 16 and 5
different compounds were considered for oxygen and iron, respectively. The set of models does not cover the whole range of possible
compounds present in the ISM. However what are commonly believed to be the main constituents (e.g. olivines, pyroxene, oxides as well as
simpler compounds) in the diffuse ISM \citep[e.g.][]{whi03,wooden08} are represented \citep[see Appendix~A, ][]{lee09}. A further effort in obtaining new measurements 
has recently been carried out
\citep[][and de Vries et al., in prep.]{lee05,lee09}. The available X-ray measurements were mostly performed on crystalline materials. The presence of dust in the form of
crystals is significantly present only in specific astronomical environments, like comets \citep{wooden08} or the inner regions of protoplanetary disks \citep{vanb04}. 
In the ISM the amount of crystalline silicates, compared to amorphous grains, is $<5$\% \citep{lidraine01}, 
therefore we do not expect to find a sizeable amount of it in our data. However, for the handful of compounds for which
X-ray laboratory measurements of both glassy and crystalline forms are available, 
we tested that the spectral difference is only appreciable if the resolution is of the order of $~3-5$\,eV. 
Therefore, with the present data, our analysis using crystalline grains is a good first order approximation of modeling the chemical composition of ID. In the glassy form, crystalline material 
loses the ordinate internal structure proper of a crystal. For simplicity, although formally not correct, we may call an amorphous silicate with a certain (for instance enstatite) stoichiometry, glassy-enstatite or glassy-MgSiO$_3$.

\subsection{Abundances and depletion}
As X-ray spectra do not display any sharp H feature, either in emission or absorption, the total hydrogen column density has been evaluated from the low energy curvature 
of the spectrum, which extends down to 36\,\AA. 
This is a reliable method, as starting already at about 25\,\AA, the transmission is largely dominated by He and H. 
The best fit value for the hydrogen column density is $N_{\rm H}=(1.63\pm0.02)\times10^{21}$\,cm$^{-2}$, 
slightly larger than the nominal values $(1.52\pm0.07)\times10^{21}$\,cm$^{-2}$ \citep{dl90} and $(1.32\pm0.05)\times10^{21}$\,cm$^{-2}$ \citep{kal05}, 
which are the average values over a region of
1$^{\circ}$ radius around the source. 
The reason could be that the spectral curvature measures the total hydrogen (i.e. \hi, \hii\ and H$_2$), 
which should be larger than \hi\ alone. The amount of H$_2$ in the diffuse ISM is relatively low \citep[e.g.][]{takei02}. 
We estimated the amount of \hii\ using the \oii/\oi\ ratio as a proxy \citep[following the expression in][]{fs71}. 
We derive that $\sim$4\% of the measured hydrogen column density is in the
form of \hii. This would reconcile our value with the \hi\ measurement mentioned above. Besides, it could be that a 
small fraction (in this case no more than $1-3\times10^{20}$\,cm$^{-2}$)  
of the total H resides in the immediate 
surroundings of the source, like in other sources \citep[e.g.][]{peter95,vanpeet}. Such a low column density would be insufficient, however, to produce deep absorption features. 
We compute the following abundances taking 
as a reference the value we measure from the X-ray spectrum.\\

In the iron edge region (Sect.~\ref{par:iron}), a fit in terms of pure gas is clearly unacceptable. 
This points easily to a further contribution from dust. According to our best fit,
the total column density of gaseous iron is then 
$N_{\rm Fe}^{\rm gas}=(0.56\pm0.08)\times10^{16}$\,cm$^{-2}$. From the iron edge fit we obtain a total dust column density of 
$N_{\rm Fe}^{\rm dust}=(3.8\pm0.5)\times10^{16}$\,cm$^{-2}$. The sum of these iron components provides  
$N_{\rm Fe}=(4.3\pm0.5)\times10^{16}$\,cm$^{-2}$. When we compare this number to 
the value predicted from our set of solar abundances, $N_{\rm Fe}^{\rm Sun}=(5.10\pm0.06)\times10^{16}$\,cm$^{-2}$ (given the best fit total hydrogen column density of 
$N_{\rm H}=(1.63\pm0.02)\times10^{21}$\,cm$^{-2}$) we obtain that the abundance is $0.85\pm0.08$ times solar. 
The depletion of iron, 
meant here as the ratio of dust over the total amount of a given element is instead $0.87\pm0.14$ (Table~\ref{t:abu}). 
This value is often reported to be higher \citep[e.g. 0.97,][]{jenkins09}. However, other studies report lower values for the Fe depletion \citep[0.7,][]{wilms00}. 
 We note that, considering some uncertainties that still remain in the edge fitting
(Sect.~\ref{par:iron}) the value we obtain may be considered as a lower limit. \\ 
At a temperature of $kT=$0.5\,eV, Fe is mainly \fei\ and only $\sim$2\% of 
the total iron in \feii. The sum of the mildly ionized phase (\feii--\feiv), excluding therefore the high ionization ions described in Sect.~\ref{par:ovii}, is $0.24\times10^{16}$\,cm$^{-2}$ along this line of sight, i.e. Fe$_{\rm ion}$/Fe$_{\rm tot}^{\rm gas}\sim$0.3, remembering that neutral iron is $\sim87$\% depleted.\\

In the oxygen region the evidence for dust is not as striking as in the iron region (Sect.~\ref{par:oxy}). From the best fit including cold ($kT=0.5$\,eV) gas, 
the amount of oxygen in the gas form is consistent to be solar ($N_{\rm O}^{\rm gas}=(9.8\pm1.0)\times10^{17}$\,cm$^{-2}$). 
However, the amount of oxygen in dust compounds is $N_{\rm O}^{\rm dust}=(2.5\pm0.2)\times10^{17}$\,cm$^{-2}$. 
The total amount of oxygen is then $N_{\rm O}=(1.2\pm0.1)\times10^{18}$\,cm$^{-2}$, showing about 23\% overabundance
 with respect to solar (Table~\ref{t:abu}). With this reference value, the amount of depletion is $0.20\pm0.02$. 
Oxygen depletion and abundances, based on \ch\ data have been previously presented by \citet{yao06}. 
As expected, the column density of \oi\ ($N_{\rm O}^{\rm gas}$) that we measure 
is similar to theirs. 
However, taking as a reference the \citet{ag89} abundances ($A_{\rm O}^{ag89}=8.5\times 10^{-4}$), \citet{yao06} find an underabundance of oxygen of about 30\%. 
In the reference list that we adopted \citep{lodders09},
the absolute oxygen abundance is significantly lower ($A_{\rm O}^{L09}=6.0\times10^{-4}$).\\
Finally, the amount of mildly ionized oxygen (\oii--\ov, with total column density $5.5\times10^{16}$\,cm$^{-2}$) 
over the total oxygen in gas form is O$_{\rm ion}$/O$_{\rm tot}^{\rm gas}\sim$0.05.\\

As shown in Sect.~\ref{par:si} the inclusion of Si in dust is 85\%, while for Mg is 97\% 
of the total ISM. These values are comparable to what has been reported in the literature 
\citep[e.g. 80--92 and 90--97\% for Mg and Si respectively,][]{wilms00,whi03}. 
However, Mg and Si could not be studied in detail in this case 
because of the relatively low column density toward \uu\ which imprints shallow absorption edges. 
Moreover, our fitting of the Mg and Si edges relies on the oxygen modeling. Therefore, if other Mg or Si compounds are present, here they are difficult to detect.  
Keeping in mind these limitations we report the depletion values of Mg and Si in Table~\ref{t:abu}, for completeness. 
The abundance estimate of both Mg and Si are again driven by the oxygen edge fitting and are formally slightly above the
solar values.\\
In Table~\ref{t:abu} we list for each element, the ratio with Proto-solar abundances and the amount of depletion defined as the ratio between dust and total ISM abundance.\\
A further test on the abundances derived from the X-ray data would of course come from high-resolution ($R=\lambda/\Delta\lambda> 20,000$) 
UV data, where the elemental gas phase can be accurately studied \citep[e.g.][]{ss96}. In the case of \uu, high-resolution UV data (either from HST or FUSE) are not available. 
\uu\ has been observed only with the HST-STIS-G140L spectrograph ($R=1,000$). 
We find that the signal to noise ratio of the data was insufficient for a quantitative study on the absorption lines.

\subsection{The location of the cold matter}
It has been established that a gradient in the abundances in our Galaxy exists for the most abundant metals. 
An average slope of 0.06\,dex\,kpc$^{-1}$ 
should roughly apply to the gradient for both O and Fe (see Sect.\ref{par:intro}). \\
In our analysis we estimate a slight overabundance of oxygen (by a factor $\sim1.2$, see Table~\ref{t:abu}), 
while iron is a factor $\sim0.85$ of the solar value. This appears to be in contradiction with the expected abundances at the distance of the source, where
according to the gradient above, some overabundance is expected.  
However, \uu\ is located at latitude $b=-7.9133^{\circ}$, i.e. about 1\,kpc below the Galactic disk, therefore our line of sight intercepts a relatively 
small fraction of the cold ISM near the source. The cold phase that we detect towards this source is rather due to absorption in the environment close to the
Sun.\\
For iron, a gradient as a function of both the radial distance from the Galactic center and the height above the disk has been estimated \citep[e.g.][based on open clusters measurements]{chen03}. Our value of
[Fe/H]\footnote{defined as log(Fe/H)--log(Fe/H)$_{\odot}$} is --0.07. This is consistent, within the errors, with absorption in the disk at the distance of the Sun rather
than absorption local to the source, at the height of \uu\ below the disk \citep{chen03}.\\
The location of the gas based on the oxygen abundance is difficult to determine, as there is a large scatter
in the measurements \citep[e.g.][ for a compilation of results]{rudolph06}. Our oxygen abundance fits well with both
absorption far or near the observer. It is therefore not straightforward to understand where the bulk of the absorption takes place on the basis of
gradient measurements. However, a systematic analysis of dust scattering halos points out that most of the scattering (and therefore the absorption by dust) should happen within $\sim 3^{\circ}$ from the
Galactic plane \citep{peter95}. This further support the intuitive idea that most of the cold absorbing material is located close to the observer.

\begin{table}
\caption{\label{t:abu}Relative abundances and depletion values from the present analysis.}
\begin{center}
\begin{tabular}{lllll}
\hline\hline
elem. &  $N^{\rm gas}$ & $N^{\rm dust}$ & $A_Z/A_Z^{\odot}$ & dust/ISM\\
\hline
N &  $13.3\pm0.1$ & 0.0 &$1.2\pm0.1$ & 0.0\\
O &  $98\pm10$ & $25\pm2$ & $1.23\pm0.03$ & $0.20\pm0.02$\\ 
Fe &  $0.56\pm0.08$ & $3.8\pm0.5$ & $0.85\pm0.08$ & $0.87\pm0.14$\\
Ne &  $20.7\pm0.2$ & 0.0 & $0.9\pm0.1$ & 0.0\\
Mg &  $<0.3$ & $7.8\pm0.8$& $1.28\pm0.13$ & $>0.97$\\
Si &  $<1.3$ & $7.8\pm0.8$ & $1.25^{+0.20}_{-0.12}$ &$>0.86$\\

\hline
\end{tabular}

\end{center}
Notes: 
Abundances are referred to \citet{lodders09}\\
Column densities are in units of $10^{16}$\,cm$^{-2}$

\end{table}

\subsection{The silicates}\label{par:silicates}
 In the present study, we find that the fit of both the iron and oxygen edges in general rejects silicate models containing iron (e.g. Fig.~\ref{f:oxy_bc},~\ref{f:fe_hot}). In the iron region, the shift of the L\,III edge is
 consistent with absorption by metallic iron. Also a modest quantity of iron in the form of \hem\ is allowed by the fit. As discussed above, in the oxygen region the evidence for
 dust is not striking. Our results may be still partially contaminated by instrumental effects, such as bad pixels in the RGS, which are however included in
 the response files using the most updated calibration. We keep limitations in mind when discussing 
 the physical implications of this result.  
 
 In summary, we find that absorption by dust is mainly caused by metallic iron and glassy-enstatite. This kind of composition is reminiscent of the composition of GEMS
 (Glass with Embedded Metal and Sulfides), that are small grains abundant among the interplanetary dust particles \citep{bradley94}. Most of 
 GEMS particles should not have an ISM origin, but they rather reside in interplanetary environments \citep{kel_mes08}. However, GEMS with anomalous composition may have been processed in the ISM
 \citep[][ and references therein]{matzel08}. In particular, some of those particles show a low amount of sulfur relative to silicon \citep[S/Si$~0.19$]{kel_mes08}, 
 which is more similar to
 what is found in the diffuse ISM \citep{sofia} rather than in the solar neighborhood \citep{ae82}. However, we cannot yet test the contribution of 
 sulfides in a typical GEMS in our data, as FeS laboratory measurements around both the S K- and the Fe L-edge are not available. 
From the model, the glassy-enstatite MgSiO$_3$ provides the best fit. Therefore the Mg/Si ratio is 1 by definition. This is also the value found for anomalous composition GEMS which are possibly of ISM origin,
\citep{kel_mes08}. Unfortunately we have not yet the means of testing compounds with a varying amount of
Mg (or Fe). Therefore, we cannot test wether a Mg/Si ratio of $\sim$0.6 \citep[typical of the average GEMS in interplanetary dust][]{kel_mes04,ishii} would be still allowed by the fit.\\
We find that the Fe/Si ratio ranges from 0.42--0.55, which matches this ratio in any type of GEMS where Fe/Si=0.43--0.54
\citep{kel_mes04,kel_mes08}.  
This is different from what is expected in the diffuse ISM \citep[Fe/Si=$0.8\pm0.2$,][]{sofia}. 
However, the Fe/Si ratio is currently under debate. Recent
studies have revised this value, lowering it in fact significantly to roughly Fe/Si$\sim0.1$ \citep[see ][ for a discussion]{min07}.\\
Finally, the amount of Fe along this line of sight is about two times less than Mg (Mg/Fe$=2.0\pm0.3$). This estimate may be biassed because we cannot tune the amount of Mg, which is bound to be equal to Si in MgSiO$_3$. 
Another caveat that should be kept in mind is that large grains would be grey to
X-ray radiation \citep{whi03}. Therefore, at least in absorption, the large grain population (which may contain more iron) can be under-represented \citep{lee09}.\\

Even taking into account the limitations imposed by our models, we find that, from both the O and Fe edges fits, Mg-rich rather than Fe-rich silicates are 
present along the line of sight to \uu. 
Such a result strengthens previous studies in both IR \citep{min07} and X-rays \citep{costantini05} which point to the same conclusion. 
Interestingly, a similar spectral analysis, but restricted to the Fe L edge only, shows that oxides, rather than olivines or pyroxene are responsible for the absorption along the line of sight
to Cyg~X-1 \citep{lee09}. 
The column density towards \uu\ is about 4.5 times smaller than for Cyg~X-1. This result may therefore point out a chemical homogeneity on different path lengths within the diffuse
ISM.

\subsection{A fast outflowing gas?}
The oxygen region of the RGS spectrum of \uu\ displays evidence, clearly detected in two set of RGS observations, of two absorption lines, 
which are consistent with 
\oiv\ and \ov\ outflowing at $v\sim1200$\,\kms (Fig.~\ref{f:oxy_best}). 
In the \uu\ accretion flow heavy elements should not be abundant, as the companion is classified as a He-white dwarf, on the basis of X-ray binary evolutionary models
\citep{1987ApJ...322..842R}. From optical spectra analysis \citet{nelemans10} found that elements heavier than Ne are absent in ultracompact systems. 
Oxygen is a relatively light element
which could still be present in the outer envelope of the companion and be transferred in the accretion disk. 
Moreover, a way of producing oxygen could be through triple-$\alpha$ burning of He in the white dwarf. Qualitatively, 
this could justify the detection of an outflow containing oxygen. However this result needs to be tested with additional observations.\\ 
Here we fitted the absorption in terms of a photoionized gas. Other lines are predicted by the model, but they are
too weak or in a noisy part of the spectrum to be significantly detected. This system is well detected in both RGS data sets, taken 8 years apart. The outflow did not change its
physical parameters during that time. We note that this models could not be tested against the \ch-HETG data, because of the
noise affecting the oxygen region. The model predicts also a substantial amount of \civ. We qualitatively checked STIS-G140L low-resolution data (taken in 1998, about 3.5 years before the first RGS measurement) 
for the presence of a blueshifted \civ\
doublet. Fixing the outflow velocity to the X-ray value we obtain an upper limit for the \civ\ column density of $<10^{13}$\,cm$^{-2}$, which is almost two orders of magnitude lower
than that required by the X-ray model. Therefore, in the hypothesis of an outflow, this must have been absent 
three and a half years before the first RGS pointing, when the STIS observation was carried out. Fast outflows are not unusual in X-ray binaries and are 
interpreted as accretion disk winds \citep[e.g.][]{jmiller06}. However, the ions involved in the outflow are generally
more highly ionized (e.g. \ovii, \neix-\nex).\\
In terms of goodness of fit, this outflowing system can be equally well fitted by a collisionally ionized plasma with temperature $T=13.9\pm0.8$\,eV and $N_{\rm
H}=3.4\pm0.5\times10^{19}$\,cm$^{-2}$. The amount of \civ\ predicted by this model would be $\sim6\times10^{14}$\,cm$^{-2}$\,cm$^{-2}$, 
less than what is predicted by the photoionized gas, but still in disagreement with the upper limit derived from the STIS data. 
However, higher-quality STIS data are necessary for a reliable comparison between the UV and X-ray band outflowing absorber.\\
Fast moving collisionally ionized clouds in the line of sight with such high velocities are not reported by the UV surveys \citep{savage04}. Therefore the
phenomenon should arise in the proximity of the source, possibly in a less ionized impact region, that would cause ionization by collisions. 
A candidate is the spot where the accretion flow impacts the accretion disk, which should be less ionized than the disk itself \citep{boirin,vanpeet}. 
The absence of emission lines from this gas implies that, regardless of the absorbing mechanism, the flow pointing toward the observer must be very collimated. 
Although plausible, the scenarios depicted above need support from further (multiwavelength) observations.\\ 
Finally, absorption by dust seems unlikely, as dust features are generally smoother (Appendix~A).

\section{Conclusions}\label{par:conclusions}
In this paper we present the X-ray analysis of the continuum of \uu\ (using a quasi-simultaneous observation of
 \xmm\ and \inte) and of the absorption features due to the cold matter in the line of sight (using \xmm-RGS and
 \ch-MEG data).\\
 
The continuum shape and the Eddington ratio show that the source was caught in a high-state. The continuum
is well fitted by black body emission plus a Comptonization component which extends up to 40\,keV. We do not find evidence of
iron emission, either from neutral or ionized matter. This may be naturally explained by the metal-poor accretion stream
expected from the white-dwarf companion.\\

The absorption spectrum shows the presence of many components with different ionization. We focused on the cold and
mildly ionized phase only. Oxygen has been found slightly overabundant by a factor 1.23 times the solar value. Iron is on the contrary 
slightly underabundant ($\sim$0.85 times solar). The abundance values are not dramatically deviating from the solar ones and do not 
allow us to assign a precise location of the absorbing gas. However, a
location close to the observer seems likely.

Thanks to the simultaneous study of absorption by dust and gas we measured also the element depletion. Oxygen is 
mildly depleted by a factor about 0.20. The
depletion of iron is more evident, as the depletion factor is 0.87. The depletion of Mg and Si are more difficult to determine. 
We find that they are depleted of a factor $>0.97$ and $>0.86$, respectively\\ 

We modeled the dust contribution with the currently
availables absorption profiles of dust compounds. Our conclusions bear the uncertainty due to the still limited dust
data-base and a lower sensitivity in selected spectral regions. 
However we clearly find that both the oxygen and iron edges cannot be fitted by
iron-rich silicates. On the contrary, the oxygen edge is consistent to be mostly absorbed by enstatite (MgSiO$_3$, 
possibly in a glass-form). Metallic iron should be the main absorber in the iron edge. This leads to the interesting
possibility that a GEMS-like form (Mg-rich silicates with metallic iron inclusion) of grain may be absorbing along this line of
sight. A fraction of the studied GEMS, in particular the sulfur-poor grains, are believed to be of ISM origin and have also been proposed as constituents of ISM. 
For the first time an X-ray absorption analysis provides a
tentative confirmation of this scenario.\\

Finally, we report the tentative detection of a mildly ionized outflow ($v_{\rm out}\sim1200$\,\kms), highlighted by the 
\oiv\ and \ov\ absorption lines. Both a photo- or collisional- ionizing process could fit the lines, leaving open the
interpretation on the nature of this gas.

\begin{acknowledgements}
 This research made use of the Chandra Transmission Grating Catalog and archive (http://tgcat.mit.edu). 
 We also made use of the Multimission Archive at the Space Telescope Science Institute (MAST). 
 STScI is operated by the Association of Universities for Research in Astronomy, Inc., under NASA contract NAS5-26555. 
 The Space Research Organization of the Netherlands is supported
financially by NWO, the Netherlands Organization for Scientific Research. \xmm\ and \inte\ are ESA science missions with instruments and contributions directly funded by ESA 
Members States and the USA (NASA). Thanks to O.~Madej, M.~Min, P.~Predehl and E.~Ratti for useful discussion. Thanks also to V.~Beckmann for pointing out the presence of 
 \inte\ data taken quasi-simultaneously to our \xmm\ data. 
\end{acknowledgements}

\begin{appendix}\label{par:appendix}
\section{Oxygen compounds}
Here we show the profiles of the absorption around the oxygen edge for the compounds used in this analysis. 
These and other profiles  are included in the AMOL model, implemented in SPEX. We refer to \citet{ciro} 
for a complete list and description of the oxygen compounds (with the exeption of MgSiO$_3$).

\begin{figure}[h]
\begin{center}
\resizebox{\hsize}{!}{\includegraphics[angle=0]{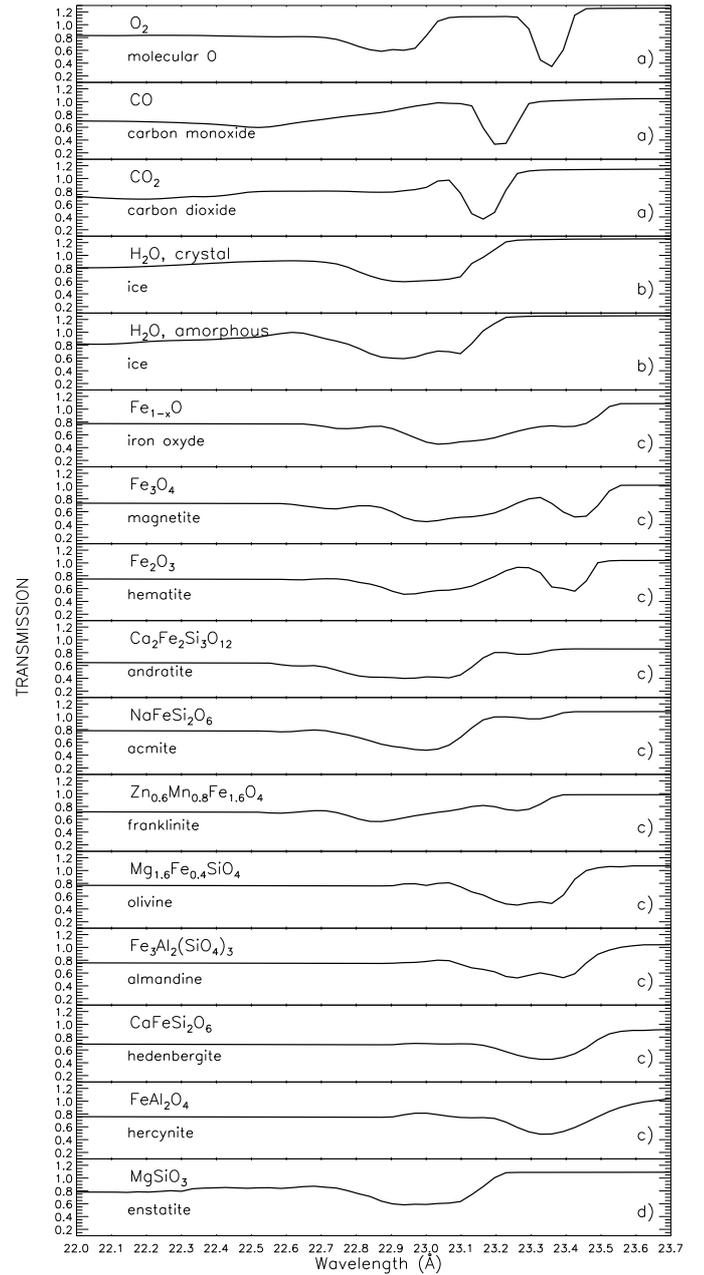}}
\end{center}
\caption{\label{f:appendix} Transmission of the dust absorption models 
included in the present analysis for the oxygen region. The oxygen column density has been set here to $10^{18}$\,cm$^{-2}$ for all compounds. 
a) \citet{barrus79}, b) \citet{parent02}, c) \citet{vaken}, d) \citet{lee08}. See also \citet{ciro} for details.}

\end{figure}

\end{appendix}

\end{document}